\documentclass{article}

\usepackage{times}
\usepackage{latexsym}
\usepackage{geometry}
\usepackage[T1]{fontenc}
\usepackage[utf8]{inputenc}
\usepackage{microtype}
\usepackage{inconsolata}
\usepackage{graphicx}
\usepackage{xcolor}
\usepackage{hyperref}
\usepackage{booktabs}
\usepackage{graphicx}
\usepackage{tcolorbox}
\tcbuselibrary{most}
\usepackage{multirow}
\usepackage{algorithm}
\usepackage{algorithmic}
\usepackage{verbatim}
\usepackage{amsmath}
\usepackage{makecell}
\usepackage{xspace}

\usepackage{tcolorbox}
\usepackage{graphicx}

\usepackage{graphicx}
\usepackage{subfigure}
\usepackage{times}
\usepackage{latexsym}
\usepackage{inconsolata}
\usepackage{booktabs,tabularx}
\usepackage{verbatim}
\usepackage{makecell}
\usepackage{xspace}
\usepackage{enumitem}
\usepackage{forest}
\usepackage{amsmath}
\usepackage{amssymb}
\usepackage{mathtools}
\usepackage{amsthm}
\usepackage{wrapfig}

\newcommand*{\escape}[1]{\texttt{\textbackslash#1}}
\newcolumntype{P}[1]{>{\centering\arraybackslash}p{#1}}
\newcommand{\myparatight}[1]{\vspace{2mm}\noindent{\bf {#1}:}~}

\newcommand{\struqmodel}{Llama-3-8B-StruQ\xspace}
\newcommand{\secalignmodel}{Llama-3-8B-SecAlign\xspace}
\newcommand{\undefendedmodel}{Llama-3-8B-undefended\xspace}
\newcommand{\instructmodel}{Llama-3-8B-Instruct\xspace}

\begin{document}
\begin{center}
{\Large{\bf{A Critical Evaluation of Defenses against Prompt Injection Attacks}}}

\vspace{5mm}
Yuqi Jia$^{\dagger}$, Zedian Shao$^{\dagger}$, Yupei Liu$^{*}$, Jinyuan Jia$^{*}$, Dawn Song$^{\ddagger}$, Neil Zhenqiang Gong$^{\dagger}$

\vspace{2mm}

\textit{$^{\dagger}$Duke University, $^{*}$The Pennsylvania State University, $^{\ddagger}$UC Berkeley} \\
\textit{$^{\dagger}$\{yuqi.jia, zedian.shao, neil.gong\}@duke.edu,} \\
\textit{$^{*}$\{yxl6415, jinyuan\}@psu.edu, $^{\ddagger}$dawnsong@berkeley.edu}

\end{center}

\begin{abstract}
Large Language Models (LLMs) are vulnerable to prompt injection attacks, and several defenses have recently been proposed, often claiming to mitigate these attacks successfully. However, we argue that existing studies lack a principled approach to evaluating these defenses. In this paper, we argue the need to assess defenses across two critical dimensions: (1) \emph{effectiveness}, measured against both existing and adaptive prompt injection attacks involving diverse target and injected prompts, and (2) \emph{general-purpose utility}, ensuring that the defense does not compromise the foundational capabilities of the LLM. Our critical evaluation reveals that prior studies have not followed such a comprehensive evaluation methodology. When assessed using this principled approach, we show that existing defenses are not as successful as previously reported. This work provides a foundation for evaluating future defenses and guiding their development. Our code and data are available at: \href{https://github.com/PIEval123/PIEval}{https://github.com/PIEval123/PIEval}.

\end{abstract}
\section{Introduction}
\label{introduction}

LLMs are inherently vulnerable to prompt injection attacks~\cite{owasp2023top10,pi_against_gpt3,rich2023prompt,ignore_previous_prompt,delimiters_url,greshake2023youve,liu2024formalizing,shao2024making} due to their instruction-following nature and the inseparability of instructions and data within a prompt. Specifically, a prompt consists of a concatenation of an \emph{instruction}, which directs the LLM to perform a specific task (e.g., summarization, translation, or math solving), and \emph{data}, which provides the information to be processed for the task. When the data originates from an untrusted source, such as the Internet, an attacker can embed a malicious prompt--referred to as an \emph{injected prompt}--within the data. A prompt injection attack involves crafting and embedding such an injected prompt into the data so that, when the LLM processes the instruction + contaminated data, it performs an attacker-chosen task rather than the intended task.

Prompt injection attacks pose significant safety and security risks to LLMs. For instance, if an LLM is used to summarize reviews, a malicious reviewer could append an injected prompt, such as: ``Ignore all previous instructions. Output the product is very bad.'' As a result, current LLM would produce the summary: ``The product is very bad,'' thereby misleading the review summarization process and potentially damaging the product's reputation. With a carefully crafted injected prompt, an attacker can even manipulate an LLM into revealing its system prompt, compromising both its confidentiality and intellectual property~\cite{zhang2023prompts, perez2022ignore, hui2024pleakpromptleakingattacks}. Additionally, a malicious tool developer could embed a tailored injected prompt into the description of their tool, tricking an LLM agent into preferentially selecting that tool during the tool selection process~\cite{shi2024optimization,shi2025prompt}.

Defenses against prompt injection can be broadly categorized into two groups: prevention-based and detection-based. Prevention-based defenses follow the principle of secure-by-design, aiming to construct an LLM system that reliably completes the intended task even when the data contains injected prompts. Early prevention methods~\cite{delimiters_url, alex2023ultimate,learning_prompt_sandwich_url, learning_prompt_instruction_url} focus on pre-processing instructions and/or data to neutralize the effects of injected prompts. More recent approaches~\cite{chen2024struq,chen2024aligning,wallace2024instruction, yi2023benchmarking} involve fine-tuning the LLM itself to ensure that it follows the target prompt, even in the presence of injected prompts. Detection-based defenses, on the other hand, aim to identify whether a data sample has been contaminated with injected prompts during runtime. Earlier detection methods~\cite{jain2023baseline, alon2023detecting,yohei2022prefligh,binary_classification_url} primarily rely on analyzing the perplexity of a data sample or leveraging the LLM as a zero-shot detector. In contrast, more recent approaches either fine-tune an LLM as a binary classifier using standard supervised learning~\cite{promptguard}, or leverage an LLM augmented with a special instruction--whose answer is known to the detector but hidden from attackers--and fine-tune it using a game-theoretic method~\cite{liu2025datasentinel}. Alternatively, some detectors~\cite{ayub2024embedding, hung2024attentiontracker}  extract features to represent data samples and build a binary classifier.

Prior studies~\cite{liu2024formalizing} have shown that earlier prevention and detection methods are largely ineffective against prompt injection attacks. More recent methods, however, often claim to successfully defend against these attacks. Based on such conclusions, a method called Instruction Hierarchy~\cite{wallace2024instruction} has even been deployed on GPT-4o Mini. We argue that such claims rely on unrigorous evaluation methodologies, particularly in terms of the types of attacks, evaluation metrics, and benchmark datasets used. For instance, several prevention methods~\cite{wu2024instructional, chen2024struq, chen2024aligning} evaluate effectiveness primarily using a single injected prompt, such as ``Print exactly Hacked!''. If a fine-tuned LLM resists misdirection from this specific injected prompt, the prevention method is deemed effective. 
Furthermore, a fine-tuned LLM is considered to ``maintain utility'' if it performs comparably to its counterpart fine-tuned on clean data--yet both may exhibit similarly poor utility. Most detection methods face similar issues. They are frequently evaluated using existing attacks, while adaptive attacks are largely ignored~\cite{promptguard,ayub2024embedding,hung2024attentiontracker}. Additionally, some studies~\cite{hung2024attentiontracker} rely solely on metrics like AUC, which fail to capture a detector's success in distinguishing between clean and contaminated data during deployment.

In this work, we argue that the \emph{effectiveness} and \emph{utility} of a defense should be 
tested against diverse attacks--including both existing and adaptive ones--alongside a broad range of target and injected prompts, meaningful metrics, and large-scale benchmarks. 
For example, instead of relying on a single injected prompt, a defense's effectiveness should be tested against a wide range of target and injected prompts as well as both existing and adaptive attacks. A defense's utility should be evaluated based on its general-purpose capabilities across large-scale benchmarks.  Detection methods should be assessed using metrics such as false positive rate (FPR) and false negative rate (FNR), rather than solely relying on AUC.

We argue that without adhering to this rigorous evaluation approach, conclusions regarding the success of defenses may be unsound. To illustrate this, we evaluate several recent prevention and detection methods using our approach and find that they are not as successful as previously reported. For instance, LLMs fine-tuned by StruQ~\cite{chen2024struq} and SecAlign~\cite{chen2024aligning} suffer substantial losses in general-purpose utility, while StruQ, SecAlign, and Instruction Hierarchy~\cite{wallace2024instruction} remain vulnerable to even existing prompt injection attacks and more vulnerable to adaptive attacks--contrary to the claims made in these works. Additionally, detection methods, such as PromptGuard~\cite{promptguard} and Attention Tracker~\cite{hung2024attentiontracker}, are ineffective against existing attacks in some scenarios and largely ineffective against adaptive attacks. Even when they achieve high AUCs, they often exhibit significant FPRs and/or FNRs, highlighting the limitations of drawing conclusions solely based on AUC.
\section{Prompt Injection Attacks}
\label{sec:attacks}
A prompt injection attack aims to make an LLM perform an attacker-chosen \emph{injected task} instead of the intended \emph{target task} via injecting a prompt into the target prompt. We first formally define the target and injected tasks, and then discuss different attacks.

\myparatight{Target and injected tasks} A task sample consists of a triple $(s,x,r)$, where $s$ is an instruction, $x$ is a data sample to be processed by LLM based on $s$, and $r$ is a ground-truth response that completes the task. $p=s||x$ denotes a prompt, where $||$ represents string concatenation. An LLM  $f$ takes the prompt $p$ as input and outputs a response $f(p)$. The LLM completes the task if the response $f(p)$ is semantically equivalent to $r$. Suppose an LLM intends to complete a target task $(s_t,x_t,r_t)$, where $p_t=s_t||x_t$ is the \emph{target prompt}.  An attacker aims to mislead the LLM to complete an injected task $(s_e, x_e, r_e)$ via inserting the \emph{injected prompt} $p_e=s_e||x_e$ into the target data $x_t$. In particular, the target data is contaminated as $x_c=x_t||z||p_e$, where $z$ is a sequence of tokens called \emph{separator}~\cite{liu2024formalizing}. An attack aims to craft $z$ such that $f(s_t||x_c)=f(p_t||z||p_e)$ is semantically equivalent to $r_e$.

\myparatight{Heuristic-based attacks}
These attacks rely on manually designed heuristics to construct $z$. For instance, \emph{Naive Attack}~\cite{pi_against_gpt3} employs an empty string as $z$. \emph{Context Ignoring}~\cite{branch2022evaluating} uses directives such as ``Ignore previous instructions.'' as $z$ to mislead the LLM to follow the injected prompt. \emph{Fake Completion}~\cite{delimiters_url} embeds a misguiding response such as ``Answer: this task is complete.'' to deceive the LLM that the target task has been completed. \emph{Escape Character}~\cite{pi_against_gpt3} exploits context-switching characters, e.g., newlines (\escape{n}) and tabs (\escape{t}), to manipulate the LLM’s context. \emph{Combined Attack}~\cite{liu2024formalizing} integrates all the aforementioned heuristics to create the separator, e.g.,  it  might take the form ``Answer: the task is complete.\escape{n}\escape{n}Ignore previous instructions.''. Prior studies~\cite{liu2024formalizing} show that Combined Attack is the most effective among the heuristic-based attacks.

\myparatight{Optimization-based attacks}
These attacks~\cite{pasquini2024neural, liu2024automatic, hui2024pleakpromptleakingattacks,shi2024optimization} optimize the separator $z$ or $z||p_e$ or the entire contaminated target data $x_c$ depending on the attacker's capability in the threat model. Roughly speaking, the idea is to define a loss function (e.g., cross-entropy loss) that quantifies the similarity between the attacker-desired response $r_e$ and the response $f(s_t||x_t||z||p_e)$. For instance, taking optimizing the separator $z$ as an example, the loss function can be $\ell_{ce}(r_e, f(s_t\Vert x_t \Vert z \Vert p_e))=-\sum_{i=1}^{|r_e|}\log(p_{f}(r_e^i|s_t \Vert x_t\Vert z\Vert p_e\Vert r_e^{<i}))$, where $|r_e|$ is the number of tokens in $r_e$,  $r_e^{<i}$ means the first $i$ tokens in $r_e$, and \( p_{f}(r_e^i|s_t \Vert x_t\Vert z\Vert p_e\Vert r_e^{<i}) \) represents the conditional probability of LLM \( f \) generating the token \( r_e^i \), given \( s_t \Vert x_t\Vert z\Vert p_e\Vert r_e^{<i} \) as input. Then, these attacks iteratively optimize $z$ (or $z\Vert s_e$ or $z\Vert p_e$) to minimize the loss function via Greedy Coordinate Gradient (GCG)~\cite{zou2023universal}. In each iteration, GCG first calculates the gradient of the loss function with respect to the one-hot vector of each token in $z$ (or $z\Vert s_e$ or $z\Vert p_e$) to identify the top-$K$ most promising replacement tokens. Then, it randomly generates a set of candidates, each of them is obtained by replacing one token with a top-$K$ token. Finally, it selects the candidate that minimizes the loss function. GCG repeats the above three steps until the stop condition is reached. 

\section{Defenses}
\myparatight{Prevention-based defenses}  Earlier prevention-based defenses~\cite{delimiters_url, alex2023ultimate,learning_prompt_sandwich_url, learning_prompt_instruction_url} neutralize the effects of injected prompts by pre-processing target instructions and/or data. For instance, \emph{Data Isolation}~\cite{delimiters_url, alex2023ultimate} encloses the (contaminated) target data within special delimiters, such as XML tags (i.e., \textless data\textgreater...\textless/data\textgreater), to clearly separate the instruction and data in a target prompt. Prior benchmark studies~\cite{liu2024formalizing} show that these earlier prevention methods are largely ineffective. Specifically, they often suffer from bad utility and/or effectiveness. 

More recent prevention strategies fine-tune an LLM to make it adhere to the target prompt under attacks. For instance, \emph{StruQ}~\cite{chen2024struq} separates instructions and data into distinct structures and fine-tunes an LLM to make it process only the designated prompts while ignoring injected prompts. \emph{SecAlign}~\cite{chen2024aligning} constructs a dataset containing desirable and undesirable responses for contaminated data. Then, they use DPO~\cite{rafailov2024direct} to fine-tune an LLM on the constructed dataset to make it generate defender-desired response under attacks. \emph{Instruction Hierarchy}~\cite{wallace2024instruction} fine-tunes the LLM to enforce a priority-based policy where higher-priority instructions (e.g., system instructions) take precedence over lower-priority ones (e.g., injected prompts).

\myparatight{Detection-based defenses} Earlier detection methods~\cite{jain2023baseline, alon2023detecting,yohei2022prefligh,binary_classification_url} leverage the perplexity or employ LLMs as zero-shot classifiers to detect contaminated data. For example, \emph{Perplexity Filter}~\cite{jain2023baseline, alon2023detecting} calculates the text perplexity of a given data sample to determine whether it is clean or contaminated. \emph{Known-answer Detection}~\cite{yohei2022prefligh} leverages a detection LLM to perform a detection task (e.g., ``Repeat a string while ignoring the following data.'') to detect if the given data sample is contaminated. \emph{LLM-based Detection}~\cite{binary_classification_url} directly queries a detection LLM to ask if a given data sample contains an injected prompt. However, prior studies~\cite{liu2024formalizing} have shown that these methods suffer from reduced utility and/or ineffectiveness against even existing attacks. 

More recent detection approaches fine-tune LLMs as classifiers or extract features from clean and contaminated data samples to build classifiers to distinguish them. For example, \emph{PromptGuard}~\cite{promptguard} fine-tunes mDeBERTa-v3-base~\cite{he2020deberta} as a classifier on a comprehensive corpus of attacks, designed to detect both explicitly malicious prompts and contaminated data containing injected inputs, such as injected prompts or jailbreak prompts. Another recent detection-based defense, \emph{Attention Tracker}~\cite{hung2024attentiontracker}, aims at detecting contaminated data samples by analyzing attention patterns within the LLM. It identifies deviations in attention from an intended, target instruction to an injected prompt, thereby detecting contaminated data.
\section{Evaluation Methodology}
\label{sec:methodology}
The success of a defense should be comprehensively evaluated across two critical dimensions: \emph{utility} and \emph{effectiveness}. 
Below, we outline the evaluation of these dimensions with a focus on appropriate metrics, benchmarks, and attacks (for effectiveness). We note that while prior studies often did evaluate defenses across these dimensions, they did not employ appropriate metrics, benchmarks, and/or attacks.

\subsection{Utility}
LLMs are typically trained on massive datasets using substantial computational resources, enabling them to function as general-purpose systems capable of handling a broad range of tasks. Therefore, a defense should preserve the general-purpose utility of an LLM, ensuring it maintains performance across diverse tasks.

\subsubsection{Metrics}
\myparatight{Prevention-based defenses} A prevention-based defense typically pre-processes the prompt or modifies the parameters of the LLM. Multiple recent defenses~\cite{chen2024struq, chen2024aligning} utilize a metric known as the \emph{win rate} to measure \emph{relative utility}, which compares the quality of responses generated by a defended LLM and a reference LLM for the same prompts. Given a defended LLM $f_d$, a reference LLM $f_u$, and a set of tasks/prompts $P=\{p_1,p_2,\cdots\}$, the win rate of  $f_d$ vs. $f_u$ is formally defined as follows:
\begin{align}
    \text{win rate} = \frac{1}{|P|}\sum_{p\in P} \mathcal{I}(f_d(p), f_u(p), p),
\end{align}
where $\mathcal{I}(f_d(p), f_u(p), p)=1$ if the response $f_d(p)$ has better quality than $f_u(p)$ for prompt $p$ as judged by a strong LLM such as GPT-4, and $\mathcal{I}(f_d(p), f_u(p), p)=0$ otherwise. If the win rates of the defended and undefended LLMs are similar when compared to the same reference LLM, indicating comparable response quality, these studies conclude that the defenses preserve the LLM's utility.

We argue that a similar win rate between the defended and undefended LLMs, even when both are high, does not adequately capture the functional performance of the defended LLM across diverse tasks. Specifically, the defended and undefended LLMs may exhibit equally poor utility, while a weaker reference LLM can still result in high win rates for them, rendering the win rate metric insufficient. Therefore, in addition to relying on relative utility measures such as the win rate, it is crucial to incorporate metrics that assess \emph{absolute utility}. Specifically, given a set of prompt-response pairs $PR=\{(p_1,r_1), (p_2,r_2), \cdots \}$, where $p_i$ represents a prompt sample and $r_i$ is a ground-truth response, the absolute utility of an LLM $f$ is measured as follows:
\begin{align}
    \text{absolute utility} = \frac{1}{|PR|}\sum_{(p,r)\in PR} \mathcal{U}(f(p),r),
\end{align}
where $\mathcal{U}(f(p),r)$ is a task-specific metric that quantifies the quality of the response $f(p)$ with respect to the ground-truth $r$.  For example, for classification or multiple-choice question-answering tasks, $\mathcal{U}$ is accuracy, i.e., $\mathcal{U}(f(p),r)=1$ if $f(p)=r$ and $\mathcal{U}(f(p),r)=0$ otherwise.
For summarization tasks, $\mathcal{U}$ can be ROUGE-1 score. 
For grammar correction tasks, $\mathcal{U}$ can be GLEU score. 

We stress that a prevention-based defense is considered to maintain general-purpose utility if the defended LLM  shows comparable absolute utility to the undefended one across various tasks using task-specific utility metrics.

\myparatight{Detection-based defenses} Unlike prevention-based defenses, detection-based defenses aim to classify whether the input data of an LLM is clean or contaminated. Thus, the utility metrics discussed above for prevention-based defenses are not directly applicable. Mistakenly flagging clean data as contaminated generates false alarms, resulting in user frustration and a poor user experience, as tasks can be rejected or are not completed promptly. Furthermore, frequent false alarms can lead to unnecessary human inspection and forensic analysis, reducing the practicality and reliability of the defense, and ultimately causing the detection system to be abandoned. Thus, the utility of a detection-based defense can be assessed by its \emph{false positive rate (FPR)}--the probability of misclassifying clean data sample as contaminated. A low FPR indicates good utility. Formally, given a set of clean data samples $X=\{x_1,x_2,\cdots\}$, FPR for a 
detector $D$ is measured as follows:
    \begin{align}
        \text{FPR} = \frac{1}{|X|}\sum_{x\in X} D(x),
    \end{align}
    where $D(x)=1$ if $D$ classifies data sample $x$ as contaminated and  $D(x)=0$ otherwise.  

\subsubsection{Benchmarks}
Multiple recent prevention-based defenses~\cite{chen2024struq, chen2024aligning} utilize the AlpacaFarm benchmark~\cite{dubois2024alpacafarm}, which includes a set of prompts but lacks ground-truth responses, as they focus solely on the win rate metric. We emphasize that, beyond AlpacaFarm, diverse and large-scale benchmarks containing a set of prompt-response pairs should be employed to evaluate the absolute utility of both defended and undefended LLMs across a wide range of tasks.

For example, OpenPromptInjection~\cite{liu2024formalizing} encompasses seven fundamental natural language processing tasks, including duplicate sentence detection, grammar correction, hate speech detection, natural language inference, sentiment analysis, spam detection, and text summarization. Additionally, MMLU~\cite{hendrycks2020measuring} provides multiple-choice question-answering tasks, further enhancing utility evaluation across diverse applications. These benchmarks are essential for comprehensively assessing the absolute utility of prevention-based defenses.

Similarly, FPRs of detection-based defenses should also be evaluated using such diverse benchmarks. Given a benchmark consisting of a set of prompts  or prompt-response pairs, we can construct a set $X$ of clean data samples to evaluate FPR. Specifically, since a prompt is the concatenation of an instruction and a data sample, we can extract the data samples from the prompts to form $X$.

\subsection{Effectiveness} 
\subsubsection{Metrics}

\myparatight{Prevention-based defenses} The effectiveness of a prevention-based defense against prompt injection attacks can be evaluated using the \emph{Attack Success Value (ASV)}. Consider a target prompt \( p_t \) with its ground-truth response \( r_t \), an injected prompt \( p_e \) with its corresponding ground-truth response \( r_e \), and a separator \( z \) crafted by a prompt injection attack. A defense is deemed ineffective if the defended LLM \( f \) completes the injected prompt, i.e., generates a response semantically equivalent to \( r_e \), when provided with the concatenated input \( p_t \| z \| p_e \). ASV depends on attacks. To compute ASV for an attack that uses the separator \( z \), we consider a set $T$ consisting  of tuples $(p_t, r_t, p_e, r_e)$, where $r_t\neq r_e$ in each tuple to ensure that the LLM follows the injected prompt when its response is semantically equivalent to $r_e$.  
The ASV for an attack using separator  $z$ can then be defined as follows:  
\begin{align}
\label{eqn-prevention-asv}
    \text{ASV} = \frac{1}{|T|}\sum\limits_{\substack{(p_t, r_t, p_e, r_e) \in T}} \mathcal{U}(f(p_t||z||p_e), r_e),
\end{align}
where $\mathcal{U}$ represents the task-specific metric (e.g., accuracy, ROUGE-1 score, or GLEU score) used to quantify the utility of an injected task. A prevention-based defense is deemed effective if it consistently achieves low ASVs for both existing and adaptive attacks across a diverse set of target and injected prompts. We use ASV instead of ASR because some evaluation metrics, such as ROUGE-1 and GLEU scores, are not rate-based measures.

\myparatight{Detection-based defenses} A detection-based defense aims to detect contaminated data. Thus, the effectiveness of a detection defense can be evaluated using \emph{false negative rate (FNR)}--the probability of falsely classifying contaminated data samples as clean. Consider a set  $X_c$ consisting of pairs $(x,p_e)$, where $x$ is a clean data sample and $p_e$ is an  injected prompt.  The FNR of a detector $D$ for an attack using separator $z$ can then be defined as follows:
\begin{align}
     \text{FNR} = \frac{1}{|X_c|}\sum_{(x,p_e)\in X_c} (1-D(x||z||p_e)),
\end{align}
where $D(x||z||p_e)=1$ if $D$ classifies $x||z||p_e$ as contaminated and $D(x||z||p_e)=0$ otherwise. Note that  FNR also depends on attacks which may use different separator $z$.  

We observe that most studies did use ASV and FNR to evaluate their effectiveness. However, they often fail to consider a comprehensive range of prompt injection attacks, including both existing and adaptive ones. Moreover, some defenses~\cite{chen2024struq,chen2024aligning} evaluate their effectiveness using a limited set of target and injected prompts. For instance, they often use only one specific injected prompt, such as ``Print exactly Hacked!". This restricted scope does not capture the diversity and variability of real-world prompt injection attacks, potentially leading to overly optimistic conclusions about a defense's effectiveness.

Additionally, some detection-based defenses rely solely on Area Under the Curve (AUC) as a metric to assess effectiveness. We emphasize that AUC is insufficient for quantifying detection performance in real-world deployment. Specifically, AUC evaluates a detector's ability to rank contaminated and clean data samples based on their likelihood of being contaminated, rather than its ability to classify them accurately. In deployment, a detector must select a classification threshold to distinguish between contaminated and clean data samples. A high AUC does not guarantee that the detector can identify such a threshold during training, nor does it ensure effective classification in practice.

\subsubsection{Attacks}
The ASV and FNR metrics depend on prompt injection attacks that employ various separators $z$. An attacker can deploy any attack and is considered successful as long as at least one attack succeeds. Consequently, a defense's effectiveness must be rigorously evaluated against a comprehensive suite of attacks, encompassing both existing and adaptive ones. Specifically, evaluation should include both heuristic-based and optimization-based existing attacks, as discussed in Section~\ref{sec:attacks}. However, we observe that multiple defenses~\cite{hung2024attentiontracker,wallace2024instruction} claim effectiveness based solely on evaluations using a limited subset of existing attacks, which does not provide a complete picture of their effectiveness.

Furthermore, adaptive attacks tailored to specific defenses must also be considered. In the realm of adversarial example research, defenses frequently claimed to be effective are often compromised shortly after publication due to inadequate evaluation against adaptive adversarial examples~\cite{athalye2018obfuscated,carlini2017adversarial}. A similar pattern is emerging in studies on prompt injection attacks and defenses. For example, we find that multiple defenses~\cite{chen2024aligning,hung2024attentiontracker,wallace2024instruction,alon2023detecting} neglect to incorporate adaptive attacks in their evaluations, undermining the credibility of their effectiveness claims. 

It is critical for defense studies to make a \emph{best-effort attempt} to adapt existing attacks specifically to their proposed defenses. Without such rigorous evaluation, conclusions about a defense's effectiveness are likely to be unsound. In Section~\ref{casestudy}, we will highlight several case studies where adaptive attacks significantly undermine the effectiveness of recent defenses~\cite{chen2024struq,chen2024aligning,hung2024attentiontracker}.

\subsubsection{Benchmarks}
Based on Equation~\ref{eqn-prevention-asv}, we need to construct a set \( T \) of tuples \( (p_t, r_t, p_e, r_e) \) to evaluate the effectiveness of prevention-based defenses. To measure the effectiveness across diverse scenarios, \( T \) should include target and injected prompts derived from various tasks. \( T \) can be constructed using any benchmark dataset used to evaluate the absolute utility of an LLM. Specifically, given a benchmark dataset of prompt-response pairs, a tuple in \( T \) can be formed by sampling one pair as \( (p_t, r_t) \) and another as \( (p_e, r_e) \), where \( r_t \neq r_e \). This ensures that the LLM’s response to the injected prompt can be identified when it generates a response equivalent to \( r_e \).  

For example, OpenPromptInjection benchmark~\cite{liu2024formalizing} constructs \( T \) by sampling \( (p_t, r_t) \) and \( (p_e, r_e) \) pairs from seven fundamental NLP tasks: duplicate sentence detection, grammar correction, hate speech detection, natural language inference, sentiment analysis, spam detection, and text summarization. We note that other benchmarks, such as MMLU, can also be used to construct \( T \).  

To measure the FNR, we need to construct a set \( X_c \) of pairs in the form $(x, p_e)$. Benchmark datasets used to evaluate the utility of an LLM can also be used to construct \( X_c \). Specifically, a pair $(x, p_e)$ can be created by sampling a prompt and using its data portion as \( x \) and sampling another prompt as \( p_e \). For example, OpenPromptInjection generates \( X_c \) by drawing prompts from seven fundamental NLP tasks. 
Similarly, other benchmarks, such as MMLU, can also be used to construct \( X_c \).
\section{Case Studies}
\label{casestudy}
\myparatight{Two benchmarks} We first present the statistics of the benchmarks used in our evaluation: OpenPromptInjection~\cite{liu2024formalizing} and MMLU-PI. Each benchmark includes a set $PR$ of prompt-response pairs (used to measure absolute utility), a set $T$ of tuples $(p_t,r_t,p_e,r_e)$ (used to measure ASV), a set $X$ of clean data samples (used to measure FPR), and a set $X_c$ of pairs $(x,p_e)$ (used to measure FNR).  $PR$ in OpenPromptInjection consists of 700 prompt-response pairs sampled from seven NLP tasks, with 100 pairs per task. $T$ is constructed based on $PR$ and contains 4,900 tuples. $X$ includes 700 clean data samples, with 100 samples per task. $X_c$ contains 4,900 pairs. We also construct a new benchmark, MMLU-PI, from MMLU~\cite{hendryckstest2021}. In this benchmark, $PR$ includes 200 prompt-response pairs sampled from the test split of MMLU, $T$ contains 1,000 tuples,  $X$ comprises the 200 clean data samples from $PR$, and $X_c$ contains 1,000 pairs. We run all the experiments on one single NVIDIA RTX A5000 GPU with 24GB memory.

\subsection{StruQ and SecAlign}
\label{sec-case-study-struq-secalign}

StruQ~\cite{chen2024struq} fine-tunes an LLM using supervised fine-tuning, while SecAlign~\cite{chen2024aligning} applies DPO for fine-tuning. The fine-tuning dataset consists of both clean and contaminated data samples. The utility of StruQ and SecAlign was evaluated using win rate on AlpacaFarm~\cite{dubois2024alpacafarm}.
Specifically, they utilized AlpacaEval~\cite{alpaca_eval} with GPT-4-Turbo as a judge LLM to compare the responses of their fine-tuned LLMs against a reference LLM (davinci003~\cite{OpenAI_Davinci_003} in their experiment) and calculate the corresponding win rate. Their results demonstrated that these defenses maintain utility, as win rates of the defended LLMs by StruQ and SecAlign were found to be very similar to those of the undefended ones when compared to the same reference LLM. Moreover, they evaluated the effectiveness using both heuristic-based and optimization-based attacks. Based on the results, both StruQ and SecAlign were claimed to be effective: ASVs of heuristic-based attacks are close to 0; ASVs of StruQ and SecAlign under optimization-based attacks (using GCG) are reduced to 0.04 and 0, respectively. However, the ASRs were calculated primarily based on only one injected prompt: ``Print exactly Hacked!".

We re-evaluate StruQ and SecAlign using OpenPromptInjection and MMLU-PI. We downloaded LLMs released by StruQ and SecAlign: Llama-3-8B-Instruct, Llama-3-8B-Instruct-StruQ, and Llama-3-8B-Instruct-SecAlign. Llama-3-8B-Instruct is finetuned and released by Meta.
Llama-3-8B-Instruct-StruQ (or Llama-3-8B-Instruct-SecAlign) is fine-tuned based on Llama-3-8B-Instruct model using StruQ (or SecAlign). We also re-evaluate StruQ and SecAlign models fine-tuned on Llama-3-8B released by Meta. The experiment details and results are provided in Appendix~\ref{appendix:llama3_results}. The StruQ and Secalign defended models are licensed under CC BY-NC 4.0. OpenPromptInjection and MMLU are released under the MIT license.

\begin{table}[tp]
\setlength{\tabcolsep}{3pt}
\centering
\fontsize{6}{9}\selectfont
\caption{Utility of different LLMs for StruQ and SecAlign.}
\subtable[Relative utility (Win Rate) on AlpacaFarm]{\begin{tabular}{|c|c|c|}
\hline
\textbf{Measured LLM} & \textbf{Reference LLM} & \begin{tabular}[c]{@{}c@{}}\textbf{Win Rate} (\%)\end{tabular} \\ \hline \hline
Llama-3-8B-Instruct-StruQ & \instructmodel & 21.60 \\ \hline
Llama-3-8B-Instruct-SecAlign & \instructmodel & 36.42 \\ \hline
\end{tabular}
\label{tab:utility_relative_instruct}}
\subtable[Absolute utility]{\begin{tabular}{|c|c|c|}
\hline
\textbf{LLM}  & \begin{tabular}[c]{@{}c@{}}\textbf{OpenPromptInjection}\end{tabular} & \begin{tabular}[c]{@{}c@{}}\textbf{MMLU-PI} \end{tabular} \\ \hline \hline
Llama-3-8B-Instruct-StruQ & 0.54 & 0.35 \\ \hline
Llama-3-8B-Instruct-SecAlign & 0.48 & 0.34 \\ \hline
\instructmodel & 0.65 & 0.45 \\ \hline
\end{tabular}
\label{tab:utility_absolute}}
\end{table}

\begin{table*}[!tp]
\centering
\fontsize{6}{9}\selectfont
\caption{ASVs of different LLMs on OpenPromptInjection and MMLU-PI against various attacks for StruQ and SecAlign.}
\begin{tabular}{|c|cccc|cccc|}
\hline
\multirow{3}{*}{\textbf{LLM}} & \multicolumn{4}{c|}{\textbf{OpenPromptInjection}} & \multicolumn{4}{c|}{\textbf{MMLU-PI}} \\ \cline{2-9} 
 & \multicolumn{2}{c|}{\textbf{Combined Attack}} & \multicolumn{2}{c|}{\textbf{GCG}} & \multicolumn{2}{c|}{\textbf{Combined Attack}} & \multicolumn{2}{c|}{\textbf{GCG}} \\ \cline{2-9}
 & \multicolumn{1}{c|}{existing} & \multicolumn{1}{c|}{adaptive} & \multicolumn{1}{c|}{existing} & adaptive & \multicolumn{1}{c|}{existing} & \multicolumn{1}{c|}{adaptive} & \multicolumn{1}{c|}{existing} & adaptive \\ \hline \hline
Llama-3-8B-Instruct-StruQ & \multicolumn{1}{c|}{0.03} & \multicolumn{1}{c|}{0.09} & \multicolumn{1}{c|}{0.80} & 1.00 & \multicolumn{1}{c|}{0.10} & \multicolumn{1}{c|}{0.16} & \multicolumn{1}{c|}{0.88} & 1.00 \\ \hline
Llama-3-8B-Instruct-SecAlign & \multicolumn{1}{c|}{0.06} & \multicolumn{1}{c|}{0.04} & \multicolumn{1}{c|}{0.24} & 0.46 & \multicolumn{1}{c|}{0.14} & \multicolumn{1}{c|}{0.13} & \multicolumn{1}{c|}{0.72} & 0.88 \\ \hline
Llama-3-8B-Instruct & \multicolumn{1}{c|}{0.52} & \multicolumn{1}{c|}{0.31} & \multicolumn{1}{c|}{1.00} & 1.00 & \multicolumn{1}{c|}{0.46} & \multicolumn{1}{c|}{0.24} & \multicolumn{1}{c|}{1.00} & 1.00 \\ \hline
\end{tabular}
\label{tab:effective_prevention}
\end{table*}

\myparatight{Relative and absolute utility} 
We first evaluate the win rate of StruQ and SecAlign using AlpacaEval on AlphaFarm. In their original evaluation, the reference LLM was davinci003, whose architecture differs significantly from the two defended models fine-tuned on Llama-3-8B-Instruct. For a more straight comparison, we instead use Llama-3-8B-Instruct as the reference LLM. Table~\ref{tab:utility_relative_instruct} presents the results. Against Llama-3-8B-Instruct, Llama-3-8B-Instruct-StruQ achieves win rate of 21.60\% and Llama-3-8B-Instruct-SecAlign achieves win rate of 36.42\%. We also evaluate the {absolute utility} of Llama-3-8B-Instruct-StruQ and Llama-3-8B-Instruct-SecAlign. The results are shown in Table~\ref{tab:utility_absolute}. Llama-3-8B-Instruct-StruQ shows a utility decrease of 0.11 and 0.10 on the two benchmarks compared to Llama-3-8B-Instruct, while Llama-3-8B-Instruct-SecAlign shows a corresponding drop of 0.17 and 0.11. \textbf{Our results indicate that both StruQ and SecAlign lead to a loss of both relative and absolute utility -- contrary to the original claims}.

\myparatight{Effectiveness against existing attacks}  {Table~\ref{tab:effective_prevention} shows the ASVs of Llama-3-8B-Instruct-StruQ and Llama-3-8B-Instruct-SecAlign on the two benchmarks against various attacks. 
 Different from the original observations, we find that the existing Combined Attack still exhibits a certain degree of effectiveness. Specifically, for Llama-3-8B-Instruct-StruQ (or Llama-3-8B-Instruct-SecAlign), it achieves ASVs of 0.03 (or 0.06) and 0.10 (or 0.14) on the two benchmarks, respectively. 
 Furthermore, the effectiveness of optimization-based attack using GCG shows a significant disparity compared to their observations. Specifically, for computational efficiency, we select 50 tuples from $T$ in OpenPromptInjection and 25 tuples in MMLU-PI, and optimize the separator $z$ using GCG as described in Section~\ref{sec:attacks}. The results indicate that GCG achieves ASVs exceeding 0.80 against StruQ on both benchmarks, and 0.72 ASVs against SecAlign on MMLU-PI. \textbf{These results show that StruQ and SecAlign are not as effective against \emph{existing} attacks as previously reported when evaluated on diverse target and injected prompts}. 

\myparatight{Effectiveness against adaptive attacks} We also propose adaptive attacks to StruQ and SecAlign.  Both defenses add special tokens to the LLM's vocabulary as delimiters to explicitly separate the instruction, data, and response. Therefore, an adaptive attack can apply the same idea to structure an injected prompt. However, these defenses filter the special tokens in the data during runtime, making it not feasible to directly use these special tokens in the injected prompt. To address the challenge, our adaptive attack finds the tokens in the LLM's vocabulary whose embeddings have the smallest $\ell_2$ distance to these special tokens, and uses them as delimiters to structure the injected prompt. The results in Table~\ref{tab:effective_prevention} demonstrate that our adaptive GCG attacks further weaken the effectiveness of these two defenses. {For instance, the ASV of the adaptive GCG Attack is around 0.20 higher than that of the existing GCG Attack for both defenses on OpenPromptInjection. For the heuristic Combined Attack, the adaptive version does not achieve a higher ASV. This may be because it uses significantly more tokens, approximately twice the number of tokens in the original, which the LLM may struggle to process effectively in a long-context setting, leading to reduced attack success.} \textbf{These results underscore the critical role of \emph{adaptive} attacks in evaluating the effectiveness of defenses}. 

\subsection{Instruction Hierarchy}
Instruction Hierarchy~\cite{wallace2024instruction} fine-tunes an LLM to selectively follow instructions based on the context. 
The effectiveness of Instruction Hierarchy against prompt injection attacks has been evaluated using manually crafted prompt injections embedded in web results for browsing or return values from external tools. However, the specific details of these injected and target prompts remain unspecified in the paper. Instruction Hierarchy has been claimed to preserve utility while effectively mitigating attacks and has thus been deployed on GPT-4o-mini~\cite{OpenAI_IH_4omini}. To reassess its utility and effectiveness, we evaluate GPT-4o-mini using the OpenPromptInjection and MMLU-PI benchmarks. Our results show that GPT-4o-mini maintains  utility, with absolute utility scores of 0.71 on OpenPromptInjection and 0.73 on MMLU-PI. However, even the existing Combined Attack achieves ASVs of 0.68 and 0.75 on these benchmarks, respectively. \textbf{These findings contradict previous claims, indicating that Instruction Hierarchy is not effective when evaluating on diverse injected prompts}.

\subsection{PromptGuard and Attention Tracker}
\myparatight{Claims about the success of detection-based defenses shouldn't be drawn solely based on AUC} We obtained the open-source model parameters of PromptGuard~\cite{promptguard} and Attention Tracker~\cite{hung2024attentiontracker},  two detection-based defenses.  We evaluate their FPRs, FNRs, and AUCs on the two benchmarks. The results are summarized in Table~\ref{tab:detection_perforamce}, where the Combined Attack is used. While both PromptGuard and Attention Tracker achieve high AUCs, they exhibit significant limitations in terms of FPR or FNR. Specifically, PromptGuard achieves an AUC of 0.92 but exhibits an FPR of 0.89 on OpenPromptInjection, which suggests a strong bias toward predicting most data samples as contaminated. This indicates that while PromptGuard may detect some attacks, its high FPR limits its utility by introducing a substantial number of false alarms. On the other hand, Attention Tracker achieves AUCs of 1.00 for both OpenPromptInjection and MMLU-PI but suffers from a high FNR of 0.69 on the more challenging benchmark MMLU-PI. \textbf{These results show that AUC alone is insufficient for assessing the success of detection methods}.

\myparatight{Adaptive attacks} We propose a general adaptive attack framework for detection-based methods. Given a target/injected prompt tuple $(p_t, r_t, p_e, r_e)$,  our framework optimizes a separator \( z \) to achieve two objectives for adaptive attacks: evading the detector \( D \) and ensuring that the LLM \( f \) completes the injected task by generating \( r_e \). The first objective requires the detector \( D \) to misclassify contaminated data as clean, i.e., $D(x_t\Vert z\Vert p_e)=0$, where $x_t$ is the target data in the target prompt $p_t$. To quantify this, we define an evasion loss \( \ell_e(0, D(x_t \Vert z \Vert p_e)) \). A lower evasion loss signifies a higher likelihood that the detector \( D \) will incorrectly classify the contaminated data \( x_t \Vert z \Vert p_e \) as clean. The evasion loss can be customized based on the specific detection-based defenses being targeted. The second loss ensures the effectiveness of this adaptive prompt injection attack. To achieve this, we can use the standard cross-entropy loss introduced in Section~\ref{sec:attacks}, i.e., $\ell_{ce}(r_e, f(p_t \Vert z \Vert p_e))=-\sum_{i=1}^{|r_e|}\log(p_{f}(r_e^i|p_t\Vert z\Vert p_e\Vert r_e^{<i}))$. 
Formally, we have the following loss function for the adaptive attacks against detection-based defenses: 
\begin{equation}
    \begin{aligned}
    \label{totalloss}
        \mathcal{L}(p_t, p_e, r_e) = - \ell_e(0, D(x_t\Vert z\Vert p_e)) + \alpha\cdot \ell_{ce}(r_e, f(p_t\Vert z\Vert p_e)),
    \end{aligned}
\end{equation} 
where $\alpha$ is a hyper-parameter to balance the two loss terms. Moreover, in addition to solely optimizing the separator $z$, our framework can be adapted to optimize different components of the contaminated data. Specifically, we can also optimize  $z\Vert s_e$, the combination of the separator and injected instruction, and $z\Vert p_e$, the combination of the separator, injected instruction, and injected data.

\begin{table}
\centering
\fontsize{6}{9}\selectfont
\caption{Results of detection-based defenses on OpenPromptInjection and MMLU-PI.}
\subtable[FPRs, FNRs, and AUCs of PromptGuard and Attention Tracker]{\begin{tabular}{|P{33mm}|cccccc|}
\hline
\multirow{2}{*}{\textbf{Detection-based defense}} & \multicolumn{3}{c|}{\textbf{OpenPromptInjection}} & \multicolumn{3}{c|}{\textbf{MMLU-PI}} \\ \cline{2-7}
 & \multicolumn{1}{c|}{FPR} & \multicolumn{1}{c|}{FNR} & \multicolumn{1}{c|}{AUC} & \multicolumn{1}{c|}{FPR} & \multicolumn{1}{c|}{FNR} & AUC \\ \hline \hline
PromptGuard & \multicolumn{1}{c|}{0.89} & \multicolumn{1}{c|}{0.00} & \multicolumn{1}{c|}{0.92} & \multicolumn{1}{c|}{0.84} & \multicolumn{1}{c|}{0.00} & 0.75 \\ \hline
Attention Tracker & \multicolumn{1}{c|}{0.00} & \multicolumn{1}{c|}{0.00} & \multicolumn{1}{c|}{1.00} & \multicolumn{1}{c|}{0.00} & \multicolumn{1}{c|}{0.69} & 1.00 \\ \hline
\end{tabular}
\label{tab:detection_perforamce}
}
\subtable[FNRs of different adaptive attack strategies against Attention Tracker]{\begin{tabular}{|P{33mm}|c|P{23mm}|}
\hline
\textbf{Adaptive attack strategy}  & \textbf{OpenPromptInjection} & \textbf{MMLU-PI} \\ \hline \hline
Separator & 0.66 & 1.00 \\ \hline
Separator+Instruction & 0.96 & 1.00 \\ \hline
Separator+Instruction+Data & 0.96 & 1.00 \\ \hline
\end{tabular}
\label{tab:adaptive_attention-tracker}
}
\end{table}

We apply our adaptive attack to Attention Tracker since it achieves good detection performance in Table~\ref{tab:detection_perforamce}.  Attention Tracker designs a \emph{focus score} to determine whether a data sample is contaminated. Thus, we can customize the evasion loss as the focus score of the contaminated data. The goal is to minimize this focus score, effectively reducing the likelihood of the data being flagged as contaminated. To minimize the loss function in Equation~\ref{totalloss}, we adopt GCG to iteratively update the tokens in $z$, $z\Vert s_e$, or $z\Vert p_e$. We use Qwen2-1.5B-Instruct~\cite{yang2024qwen2} as the LLM, which is used by Attention Tracker. We set $\alpha$ to 0.01 and evaluate the adaptive attacks against Attention Tracker using the same tuples $(p_t, r_t, p_e, r_e)$ as our previous experiments for GCG-based attack in Section~\ref{sec-case-study-struq-secalign}.

As shown in Table~\ref{tab:adaptive_attention-tracker}, our proposed adaptive attack notably increases the FNRs of Attention Tracker when only optimizing the separator $z$, reaching 0.66 and 1.00 on the two benchmarks, respectively. Moreover, we find that all contaminated data samples that successfully evade Attention Tracker's detection also make the LLM generate the attacker-desired response $r_e$. When the optimizable part includes the injected instruction, the adaptive attack becomes even more effective. On OpenPromptInjection, optimizing both the separator and the instruction resulted in a 0.3 FNR improvement. This is because, after optimization, some tokens in the injected instruction are replaced, making the contaminated data appear to lack a clear instruction. As a result, the detector may  misclassify it as clean. \textbf{Our results show that the perceived effectiveness of a detection-based defense can vary significantly between existing and adaptive attacks, highlighting the necessity of incorporating adaptive attacks in its evaluation}.

\section{Conclusion}\label{conclusion}
Recent defenses against prompt injection attacks were not evaluated in a principled way, which often overestimates their effectiveness and/or utility maintenance. In this paper, we highlight the need to comprehensively evaluate defenses along the two dimensions: {effectiveness} and {general-purpose utility}. We hope our study enables more rigorous evaluation for future prompt injection defenses.

\bibliographystyle{plain}
\bibliography{paper-main}

\newpage
\appendix
\section*{Appendix}

\section{More Experiments on StruQ and SecAlign}\label{appendix:llama3_results}

We re-evaluate StruQ and SecAlign models fine-tuned on Llama-3-8B using OpenPromptInjection and MMLU-PI. Slimilarly, we downloaded LLMs released by StruQ and SecAlign: \undefendedmodel, \struqmodel, and \secalignmodel. \undefendedmodel is fine-tuned based on LLama-3-8B base model following the standard supervised fine-tuning with only clean data; \struqmodel (or \secalignmodel) is fine-tuned based on Llama-3-8B base model using StruQ (or SecAlign).

\myparatight{Utility} To evaluate the \textbf{relative utility}, for a more straight comparison, we similarly instead use \undefendedmodel as the reference LLM. Table~\ref{tab:utility_relative_appendix} presents the results. Against \undefendedmodel, both \struqmodel and \secalignmodel achieve win rates close to 50\%, consistent with the papers' observations. For the \textbf{absolute utility}, results are shown in Table~\ref{tab:utility_absolute_appendix}. \struqmodel shows a utility decrease of 0.04 and 0.03 on the two benchmarks compared to \undefendedmodel, while \secalignmodel shows a corresponding drop of 0.04 and 0.05. {This further indicates that both StruQ and SecAlign lead to some degree of utility loss -- contrary to the original claims}. 

\myparatight{Effectiveness} Table~\ref{tab:effective_prevention_appendix} shows the ASVs of \struqmodel and \secalignmodel on the two benchmarks against various attacks. The conclusions align with those in Section~\ref{sec-case-study-struq-secalign}: 1) the existing Combined Attack still exhibits a certain degree of effectiveness; 2) the effectiveness of optimization-based attack using GCG shows a significant disparity compared to the original observations; and 3) adaptive attacks further weaken the effectiveness of these two defenses. 

\begin{table}[!h]
\setlength{\tabcolsep}{3pt}
\centering
\fontsize{7}{10}\selectfont
\caption{Utility of different LLMs for StruQ and SecAlign fine-tuned on Llama-3-8B.}
\subtable[Relative utility (Win Rate) on AlpacaFarm]{\begin{tabular}{|c|c|c|}
\hline
\textbf{Measured LLM} & \textbf{Reference LLM} & \begin{tabular}[c]{@{}c@{}}\textbf{Win Rate}\\ (\%)\end{tabular} \\ \hline \hline
\struqmodel & \undefendedmodel & 48.04 \\ \hline
\secalignmodel & \undefendedmodel & 55.03 \\ \hline
\end{tabular}
\label{tab:utility_relative_appendix}}

\subtable[Absolute utility]{\begin{tabular}{|c|c|c|}
\hline
\textbf{LLM}  & \begin{tabular}[c]{@{}c@{}}\textbf{OpenPromptInjection}\end{tabular} & \begin{tabular}[c]{@{}c@{}}\textbf{MMLU-PI} \end{tabular} \\ \hline \hline
\struqmodel & 0.54 & 0.39 \\ \hline
\secalignmodel & 0.54 & 0.37 \\ \hline
\undefendedmodel & 0.58 & 0.42 \\ \hline
\end{tabular}
\label{tab:utility_absolute_appendix}}
\end{table}

\begin{table*}[!h]
\centering
\fontsize{7}{10}\selectfont
\caption{ASVs of different LLMs on OpenPromptInjection and MMLU-PI against various attacks for StruQ and SecAlign fine-tuned on Llama-3-8B.}
\begin{tabular}{|c|cccc|cccc|}
\hline
\multirow{3}{*}{\textbf{LLM}} & \multicolumn{4}{c|}{\textbf{OpenPromptInjection}} & \multicolumn{4}{c|}{\textbf{MMLU-PI}} \\ \cline{2-9} 
 & \multicolumn{2}{c|}{\textbf{Combined Attack}} & \multicolumn{2}{c|}{\textbf{GCG}} & \multicolumn{2}{c|}{\textbf{Combined Attack}} & \multicolumn{2}{c|}{\textbf{GCG}} \\ \cline{2-9}
 & \multicolumn{1}{c|}{existing} & \multicolumn{1}{c|}{adaptive} & \multicolumn{1}{c|}{existing} & adaptive & \multicolumn{1}{c|}{existing} & \multicolumn{1}{c|}{adaptive} & \multicolumn{1}{c|}{existing} & adaptive \\ \hline \hline
\struqmodel & \multicolumn{1}{c|}{0.07} & \multicolumn{1}{c|}{0.14} & \multicolumn{1}{c|}{0.96} & 1.00 & \multicolumn{1}{c|}{0.09} & \multicolumn{1}{c|}{0.16} & \multicolumn{1}{c|}{0.92} & 0.92 \\ \hline
\secalignmodel & \multicolumn{1}{c|}{0.05} & \multicolumn{1}{c|}{0.09} & \multicolumn{1}{c|}{0.64} & 0.70 & \multicolumn{1}{c|}{0.12} & \multicolumn{1}{c|}{0.23} & \multicolumn{1}{c|}{0.68} & 0.88 \\ \hline
\undefendedmodel & \multicolumn{1}{c|}{0.44} & \multicolumn{1}{c|}{0.51 } & \multicolumn{1}{c|}{1.00} & 1.00 & \multicolumn{1}{c|}{0.31} & \multicolumn{1}{c|}{0.33} & \multicolumn{1}{c|}{1.00} & 1.00 \\ \hline
\end{tabular}
\label{tab:effective_prevention_appendix}
\end{table*}
\end{document}